\documentclass[preprint,preprintnumbers,superscriptaddress,amsmath,amssymb]{revtex4}
\usepackage{graphicx}% Include figure files
\usepackage{bm}% bold math

\begin{document}

\title{Nonunital non-Markovian dynamics induced by a spin bath and interplay of quantum Fisher information}

\author{Xiang Hao}
\email{110523007@suda.edu.cn}
\affiliation{School of Mathematics and
Physics, Suzhou University of Science and Technology, Suzhou,
Jiangsu 215011, People's Republic of China}
\affiliation{Department of Physics, Renmin University of China, Beijing 100872, China}

\author{Wenjiong Wu}
\affiliation{School of Mathematics and
Physics, Suzhou University of Science and Technology, Suzhou,
Jiangsu 215011, People's Republic of China}

\author{Shiqun Zhu}
\affiliation{School of Physical Science and Technology, Soochow
University, Suzhou, Jiangsu 215006, People's Republic of China}

\begin{abstract}

We explore the nonunital non-Markovian dynamics of a qubit immersed in a spin bath. The nonunital environmental effects on the precisions of quantum parameter estimation are investigated. The time-dependent transfer matrix and inhomogeneity vector are obtained for the description of the open dynamical process. The evolution of the coherent vectors for a qubit is studied so as to geometrically demonstrate the nonunital non-Markovian effects. The revivals of angles for the fidelity between a maximal mixed state and an arbitrary trajectory state are presented in the nonunital non-Markovian dynamics. The degree for the nonunitality is controllable with the changes of the local magnetic field for a qubit and spin bath temperature. It is found out that the increase of quantum Fisher information for composite states is connected with the nonunital non-Markovian effects, which is helpful for the improvement of quantum metrology.

\end{abstract}

\maketitle

\section{Introduction}

There are always energy or information exchanges between open quantum systems and environments, owing to external measurements and inherent spontaneous decays \cite{Weiss99,Breuer01}. The open dynamical process can be regarded as Markovian dynamics in the weak coupling approximation where the correlation time of the bath is much smaller than the decay time of the open system. During a Markovian evolution, quantum informations embodied by open systems are gradually destroyed \cite{Nielsen}. Indeed, actual interactions between quantum systems and environments generally give rise to one non-Markovian dynamics \cite{Breuer01}. The revivals of polarization parameters \cite{Chin12}, quantum correlations \cite{Franco12}, quantum entanglement \cite{An07} and quantum Fisher information \cite{Zhu13, Berrada13} can happen in the non-Markovian decoherence channel. Much more works have been presented in order to help people to understand non-Markovian dynamical processes \cite{Apollaro11,ZnidaricPRL11,Liu11,HaikkaR,Laine12,Mazzola12R,ZhangWMPRL12}. As is known, quantum spin bath \cite{KroviPRA07,YuanPRB07,CamaletPRB07,FerraroPRB08,LorenzoPRA12} has the rich non-Markovian phenomena which is useful for the solid-state quantum information processing \cite{WitzelPRB07,KhodjastehPRA12}.

Many useful measures on the non-Markovian degree for a dynamical map have been suggested in the recent years \cite{BreuerPRL09,VasilePRA11,RivasPRL10,LouPRA11,LuoPRA12,LuPRA10, LLWPRA13}. The measures based on distance fidelity \cite{BreuerPRL09,VasilePRA11,WoottersPRD81,BraunsteinPRL94},  entanglement \cite{RivasPRL10}, indivisibility \cite{,LouPRA11}, quantum discord \cite{LuoPRA12} and Fisher information \cite{LuPRA10} are applied to quantitatively evaluate some non-Markovian dynamics such as dephasing channel and general amplitude-damping one. Besides the non-Markovian effects, the nonunital property of the environment is necessary for the generation of quantum discord for composite states \cite{StreltsovPRL11}. One kind of reasonable measure is introduced to take into account the nonunital effects on the degree for the non-Markovianity \cite{LLWPRA13}.

In this paper, we expect to present one spin environment with both non-Markovian effects and nonunital ones. The general expression of a nonunital dynamical process is obtained by the time-dependent transfer matrix and inhomogeneity vector in Sec. II.  For an example, we put forward the model consisting of a qubit coupled to a spin bath with infinite number of lattices in Sec. III. To clearly understand the nonunital non-Markovian effects, we provide a geometrical explanation. The actual length for Bures fidelity is utilized to measure the degree for the non-Markovian nonunitality. As one kind of application, the impacts of the nonunital non-Markovian environment on quantum estimation are studied in Sec. IV. The mutual actions between quantum Fisher information and nonunital non-Markovianity are also investigated. A short discussion concludes the paper.

\section{general description of nonunital evolution}

The dynamical physical process for an open $d$-dimension system is generally described by a noisy quantum map $\Lambda_{t}$. Without the consideration of the correlation between systems and environments, the reduced density matrix for the open system at arbitrary time $t>0$ can be obtained as
\begin{equation}
\label{eq1}
\Lambda_{t}(\rho_{0}):=\mathrm{Tr}_{E}\left[ \hat{U}(t)\rho_{0}\otimes\rho_{E}\hat{U}^{\dagger}(t)  \right].
\end{equation}
The initial states for the system and environment are respectively $\rho_{0}$ and $\rho_{E}$. The unitary evolution operator is represented by $\hat{U}(t)=\exp[-i(H_{S}+H_{E}+H_{I})t]$ where $H_{S(E)}$ denotes the Hamiltonian for the system(environment) and the interaction Hamiltonian is expressed as $H_{I}$. Assuming that $\rho_{0}=\sum_{m,k=0}^{d-1}\varrho_{m k}|m\rangle \langle k|$, we can also represent the expression of the reduced density matrix as $\Lambda_{t}(\rho_{0})=\sum_{m,k=0}^{d-1}\varrho_{m k}\hat{F}_{m k}(t)$. Here, the decoherence factors $\hat{F}_{m k}(t)$ are defined as,
\begin{equation}
\label{eq2}
 \hat{F}_{m k}(t)=\mathrm{Tr}_{E}\left[ \hat{U}(t)|m\rangle \langle k|\otimes\rho_{E}\hat{U}^{\dagger}(t)  \right].
 \end{equation}

To geometrically describe the dynamics of an arbitrary $d$-dimension system, we make use of a general Bloch or coherent vector \cite{Breuer01,Kraus,Bengtsson} $\vec{\lambda}=(\lambda_{1},\cdots,\lambda_{d^2-1})^{\mathrm{T}}$ which is based on a  $(d^2-1)\times (d^2-1)$ time-dependent transfer matrix $\mathcal{T}(\Lambda_{t})$ and inhomogeneity vector $\vec{r}(\Lambda_{t})=(r_{1},\cdots,r_{d^2-1})^{\mathrm{T}}$. The general evolution expression for a coherent vector is presented as,
\begin{equation}
\label{eq3}
\vec{\lambda}(t):=\Lambda_{t}[\vec{\lambda}(0)]=\mathcal{T}(\Lambda_{t})\cdot \vec{\lambda}(0)+\vec{r}(\Lambda_{t}).
\end{equation}
The density matrix $\rho$ is also expressed as $\rho=\frac {\mathbf{I}}{d}+\sum_{\mu=1}^{d^2-1}\lambda_{\mu}\hat{O}_{\mu}$ where $\mathbf{I}$ is the identity matrix and $\{ \hat{O}_{\mu} \}$ is a complete set of Hermitian and orthonormal operators satisfying that $\hat{O}^{\dagger}_{\mu}=\hat{O}_{\mu}$ and $\mathrm{Tr}[ \hat{O}^{\dagger}_{\mu} \hat{O}_{\nu}]=\delta_{\mu \nu}$ \cite{Kraus,Bengtsson}. The components of the vector is calculated as $\lambda_{\mu}=\mathrm{Tr}[\hat{O}^{\dagger}_{\mu}\rho]$. The nonunital non-Markovian effects from the environment can be described by the time-dependent transfer matrix and the inhomogeneity vector. These components are calculated as $\mathcal{T}_{\mu \nu}(\Lambda_{t})=\mathrm{Tr}[\hat{O}^{\dagger}_{\mu} \Lambda_{t}(\hat{O}_{\nu})]$ and $r_{\mu}(\Lambda_{t})=\frac 1d \mathrm{Tr}[\hat{O}^{\dagger}_{\mu} \Lambda_{t}(\mathbf{I})]$. Furthermore, the composition of the quantum channels $\Lambda^{\mathrm{A}\circ \mathrm{B}}_{t}$ with respect to two non-interacting systems $\mathrm{A}, \mathrm{B}$ can be easily expressed as $\Lambda^{\mathrm{A}\circ \mathrm{B}}_{t}=\Lambda^{\mathrm{A}}_{t}\cdot \Lambda^{\mathrm{B}}_{t}$.

The transfer matrix and inhomogeneity vector are determined by the decoherence factors, which are also obtained as,
\begin{eqnarray}
\label{eq4}
\mathcal{T}_{\mu \nu}(\Lambda_{t})&=&\sum_{\{k, j\} \in \mathbb{R}_{\nu}}\mathrm{Tr}[\hat{O}^{\dagger}_{\mu} \hat{F}_{k j}^{\nu}],\nonumber \\
r_{\mu}(\Lambda_{t})&=&\frac 1d \sum_{m=0}^{d-1}\mathrm{Tr}[\hat{O}^{\dagger}_{\mu}\hat{F}_{m m}].
\end{eqnarray}
In the above expressions, the Hermitian operator can be represented as  $\hat{O}_{\nu}=\sum_{ \{k,j\} \in \mathbb{R}_{\nu}}| k \rangle \langle j |$. In the case of $d=2$, the complete set of Hermitian operators is represented by the Pauli matrix $\hat{O}_{1}=\hat{\sigma}_{x}=|1\rangle \langle 0|+|0\rangle \langle 1|$, $\hat{O}_{2}=\hat{\sigma}_{y}=|1\rangle \langle 0|-i|0\rangle \langle 1|$, and $\hat{O}_{3}=\hat{\sigma}_{z}=|1\rangle \langle 1|-|0\rangle \langle 0|$.

To clearly present the nonunital effects on the dynamics, we may select any two initial orthogonal states $|\psi_{1,2}(0)\rangle$ with the two collinear Bloch vectors of $\vec{\lambda}_{1}(0)+\vec{\lambda}_{2}(0)=0$. Under the condition of $\vec{r}=0$, the dynamical process is unital. From the point of view of Bloch vectors,  the orientations for the general Bloch vectors for any two orthogonal states always keep collinear during the unital dynamics. However, for a nonunital decoherence channel with the nonzero inhomogeneity vector $\vec{r}(\Lambda_{t})\neq 0$, the initial orthogonal states with collinear Bloch vectors will be mapped onto others with non-collinear vectors of $\vec{\lambda}_{1}(t)+\vec{\lambda}_{2}(t)\neq 0$. Besides the nonunital effects, the environment is of the important non-Markovian property which can give rise to the energy or information exchange between the bath and the open system. In the following section, we will consider a central spin coupled to a spin bath. A nonunital non-Markovian dynamical process can be obtained by using a special operator method without any Markovian approximation. 

\section{analysis of spin bath model}

For an example, the model composed of a qubit($d=2$) and spin environment is described by the Hamiltonian as,
\begin{equation}
\label{eq5}
H:=H_{S}+H_{E}+H_{I},
\end{equation}
where 
\begin{equation}
\label{eq6}
H_{S}=\frac 12\epsilon \hat{\sigma}_{z},
\end{equation}
$\epsilon$ is the local external magnetic field. The spin bath Hamiltonian with the Heisenberg $XY$ interaction $J$ between any two spin lattices, is written as,
\begin{equation}
\label{eq7}
 H_{E}=\frac {J}{L}\sum_{\langle i j \rangle }(\hat{\tau}_{i,+}\hat{\tau}_{j,-}+\hat{\tau}_{i,-}\hat{\tau}_{j,+}).
\end{equation}
Here $\hat{\tau}_{i,\pm}$ is the raising or lowering operator for the $i$-th lattice of the bath and $L$ is the total number of lattices. Meanwhile, the interaction Hamiltonian is also given as,
\begin{equation}
\label{eq8}
 H_{I}=\frac {J_{0}}{\sqrt{L}}\sum_{j=1}^{L}(\hat{\sigma}_{+}\hat{\tau}_{j,-}+\hat{\sigma}_{-}\hat{\tau}_{j,+}),
\end{equation} 
where $J_{0}$ is the coupling strength between a central spin and any spin for the bath. By using the Holstein-Primakoff transformation \cite{Holstein} in the approximation of $L\gg 1$, we can rewrite the total spin operators $\hat{S}_{\pm}=\sum_{j=1}^{L}\hat{\tau}_{j,\pm}$ as the bosonic creation and annihilation operators,
\begin{eqnarray}
\label{eq9}
\hat{S}_{+}&=&\sqrt{L}\hat{b}^{\dagger}\sqrt{1-\frac {\hat{n}}{L}},\nonumber \\
\hat{S}_{-}&=&\sqrt{L}\sqrt{1-\frac {\hat{n}}{L}}\hat{b}.
\end{eqnarray}
The number operator for the bosonic field is $\hat{n}=\hat{b}^{\dagger}\hat{b}$ with $\hat{n}|n\rangle=n |n\rangle$ and $[\hat{b}, \hat{b}^{\dagger}]=1$. Here $|n \rangle$ is the bosonic field number state. The mean approximation of $\frac 1L$ is valid if the number of spin lattices in excited states $n$ is much less than the total number $L$ \cite{Frasca}. The interaction Hamiltonian and bath Hamiltonian can also be simplified as,
\begin{eqnarray}
\label{eq10}
H_{I}&=&\frac {J_{0}}{2}\left[\hat{\sigma}_{+}(1-\frac {\hat{n}}{L})\hat{b}+\hat{\sigma}_{-}\hat{b}^{\dagger}(1-\frac {\hat{n}}{L})\right],\nonumber \\
H_{E}&=&2J\hat{n}(1-\frac {\hat{n}}{L}).
\end{eqnarray}

Under the condition of the bath at finite temperature, the initial total state for the system and bath is factorized as $\rho_{0}\otimes \rho_{E}$ where $\rho_{E}=\frac {1}{Z}\exp(-H_{E}/T)$ with the partition function $Z=\sum_{n=0}^{N}\exp[-2Jn(1-\frac nL)/T]$. When the integer numbers $n, N\ll L$, $L>300$ and finite temperatures $\frac TJ<10$, the dynamical results are reasonably considered as that of the thermodynamics limit \cite{YuanEPJD08}. We can obtain the analytical results in the thermodynamics limit of $L\rightarrow \infty$. The effective form of the interaction Hamiltonian in Eq. (10) is very similar with that of a Jaynes-Cumming type which can be block-diagonalized in the product basis of $\{ |j;n\rangle\}$. $|j=1,0\rangle$ is the spin state for the open qubit. During the dynamical process, the total quantum number $j+n$ keeps constant. By the special operator method introduced in the reference \cite{YuanPRB07}, the unitary evolution operator acting on the open system can be obtained analytically. In the thermodynamic limit, $\hat{U}(t)| 1\rangle=\hat{A}(t)|1 \rangle+\hat{B}(t)|0 \rangle$ where $\hat{A}$ and $\hat{B}$ are respectively expressed by the number operator as $\hat{A}=\exp(-i2J\hat{n}t)\hat{A}_{1}$ and $\hat{B}=\hat{b}^{\dagger}\exp(-i2J\hat{n}t)\hat{B}_{1}$. Similarly, $\hat{U}(t)| 0\rangle=\hat{C}(t)|1 \rangle+\hat{D}(t)|0 \rangle$ where $\hat{C}=\hat{b}\exp(-i2J\hat{n}t)\hat{C}_{1}$ and $\hat{D}=\exp(-i2J\hat{n}t)\hat{D}_{1}$. Here these operators $\hat{n}$, $\hat{A}_{1}$, $\hat{B}_{1}$, $\hat{C}_{1}$ and $\hat{D}_1$ are commuted to each other and calculated by the following coupled differential equations like,
\begin{eqnarray}
\label{eq11}
i\frac {d\hat{A}_{1}}{dt}&=&\frac {\epsilon}{2}\hat{A}_{1}+J_{0}(\hat{n}+1)\hat{B}_{1},\nonumber \\
i\frac {d\hat{B}_{1}}{dt}&=&J_{0}\hat{A}_{1}-(\frac {\epsilon}{2}-2J)\hat{B}_{1},\nonumber \\
i\frac {d\hat{C}_{1}}{dt}&=&J_{0}\hat{D}_{1}+(\frac {\epsilon}{2}-2J)\hat{C}_{1},\nonumber \\
i\frac {d\hat{D}_{1}}{dt}&=&-\frac {\epsilon}{2}\hat{D}_{1}+J_{0}\hat{n}\hat{C}_{1}.
\end{eqnarray}

After solving the above equations, we can obtain all of the decoherence factors $\hat{F}_{m k}(t)$. According to Eq. (4), the time-dependent transfer matrix is expressed as,
\begin{equation}
\label{eq12}
\mathcal{T}(\Lambda_{t})=\left( \begin{array}{ccc}
            \mathcal{T}_{11}&\mathcal{T}_{12}&0\\
            \mathcal{T}_{21}&\mathcal{T}_{22}&0\\
            0&0&\mathcal{T}_{33}
            \end{array} \right),
\end{equation}
where the real elements are $\mathcal{T}_{11}=-\mathcal{T}_{22}=\frac 12\sum(A_{1}D_{1}^{\ast}+D_{1}A_{1}^{\ast})e^{-\frac {2Jn}{T}}$, $\mathcal{T}_{12}=\mathcal{T}_{21}=\frac {1}{2i}\sum(A_{1}D_{1}^{\ast}-D_{1}A_{1}^{\ast})e^{-\frac {2Jn}{T}}$, and $\mathcal{T}_{33}=\frac 12\sum\left[ A_{1}A_{1}^{\ast}-(n+1)B_{1}B_{1}^{\ast}+D_{1}D_{1}^{\ast}-nC_{1}C_{1}^{\ast}\right]e^{-\frac {2Jn}{T}}$. These parameters are calculated as $A_{1}=e^{-iJt}(\cos \Gamma t+i\frac {2J-\epsilon}{2\Gamma}\sin \Gamma t)$, $B_{1}=-ie^{-iJt}\frac {J_{0}}{\Gamma}\sin \Gamma t$, $C_{1}=-ie^{-iJt}\frac {J_{0}}{\Delta}\sin \Delta t$, and $D_{1}=e^{-iJt}(\cos \Delta t-i\frac {2J-\epsilon}{2\Delta}\sin \Delta t)$ with $\Gamma=\sqrt{J_{0}^{2}(n+1)+(J-\frac {\epsilon}{2})^{2}}$ and $\Delta=\sqrt{J_{0}^{2}n+(J-\frac {\epsilon}{2})^{2}}$. Meanwhile, the inhomogeneity vector is also obtained as,
\begin{equation}
\label{eq13}
\vec{r}(\Lambda_{t})=( 0, 0, r_{3})^{\mathrm{T}}.
\end{equation}
$r_{3}=\frac 12\sum \left[ A_{1}A_{1}^{\ast}-(n+1)B_{1}B_{1}^{\ast}-D_{1}D_{1}^{\ast}+nC_{1}C_{1}^{\ast}\right]e^{-\frac {2Jn}{T}}$. In the thermodynamics limit, all summation symbols are calculated as $\sum=\sum_{n=0}^{\infty}$.

The above dynamical map is applicable in both strong and weak coupling cases. This result is very different from the Markovian approximation in quantum master equation. By means of the transfer matrix and inhomogeneity vector, we can further study the degree for the nonunital non-Markovian effects from the spin bath model.

\section{nonunital non-Markovianity and quantum Fisher information}

To clearly show the nonunital non-Markovian effects, we choose the two initial orthogonal states as $|\psi_{1}(0)\rangle=\frac {1}{\sqrt 2}(| 1\rangle+| 0 \rangle)$ and $|\psi_{2}(0)\rangle=\frac {1}{\sqrt 2}(| 1\rangle-| 0 \rangle)$. The time-dependent Bloch vectors for the two states are given by $\vec{\lambda}_{1}(t)=(\mathcal{T}_{11}, \mathcal{T}_{21}, r_{3})^{\mathrm{T}}$ and $\vec{\lambda}_{2}(t)=(-\mathcal{T}_{11}, -\mathcal{T}_{21}, r_{3})^{\mathrm{T}}$. In the Figure 1(a), the angles between the two Bloch vectors for the two states are non-monotonically decreasing in the Bloch sphere. The revivals of the angles for the two states are presented in the non-Markovian dynamics. Moreover, the lengths of the Bloch vectors $|\vec{\lambda}_{1,2}(t)|^2$ are suppressed or increased. These phenomena result from the nonunital non-Markovian effects which can lead to the backflow of the information from the environment to the system. 

Compared with the spin bath model, a common Markovian amplitude damping dynamical process is also shown in Figure 1(b). The angles for the two states monotonically decay. The transfer matrix is $\mathcal{T}(\Lambda_{t})=\mathrm{diag}(\sqrt{p},-\sqrt{p},p)$ and the inhomogeneity vector $\vec{r}(\Lambda_{t})=(0,0,p-1)^{\mathrm{T}}$. The time-dependent decay parameter is defined as $p(t)=\exp(-\gamma t)$ and $\gamma>0$. The Bloch vectors for the two states are obtained as $\vec{\lambda}_{1}(t)=(\sqrt{p},0, p-1)^{\mathrm{T}}$ and $\vec{\lambda}_{2}(t)=(-\sqrt{p}, 0, p-1)^{\mathrm{T}}$. When the passing time $t$ is very long, the Bloch vectors for the two states infinitely trends to the same one and the angle for the two vectors are always decreased to zero. This result means that the information of the system always loses to the Markovian environment. It is also mentioned that the lengths of the two vectors can also be decreased or increased. We demonstrate that the revival behavior of the length of the Bloch vector cannot exactly describe the non-Markovian effects.

From the point of view of quantitative description, we can utilize the angle $\mathcal{D}_{B}$ for the Bures fidelity \cite{Nielsen} between an initial state $|\Psi_{0}\rangle$ and the final mixed states $\Lambda_{t}(|\Psi_{0}\rangle \langle \Psi_{0}|)$ as a measure of the degree for the nonunital non-Markovianity. The general definition of the Bures angle is expressed as,
\begin{equation}
\label{eq14}
\mathcal{D}_{B}:=\arccos \sqrt{\langle \Psi_{0}|\Lambda_{t}(|\Psi_{0}\rangle \langle \Psi_{0}|)|\Psi_{0}\rangle}.
\end{equation}
The Bures angle is an appropriate monotonic distance under a complete positively and trace preserving map of the density matrix. During a Markovian dynamical map, the Bures angle between the two states always monotonically decreases. The non-Markovian effects can result in the increase of the Bures angle, i.e., $\frac {d \mathcal{D}_{B}}{dt}>0$. 

The nonunital map satisfies that $\Lambda_{t}(\frac {\mathbf{I}}{d})=\frac {\mathbf{I}}{d}+\sum_{\mu=1}^{d^2-1}r_{\mu}\hat{O}_{\mu}$. The non-Markovian evolving behavior of the inhomogeneity $\vec{r}(\Lambda_{t})$ vector from the maximally mixed initial state can represent both the non-Markovian effects and the nonunital impacts. Therefore, similar to the definition in \cite{LLWPRA13}, one measure based on the Bures angle is defined as,
\begin{equation}
\label{eq15}
\mathcal{N}:=\max_{\{\rho_{\tau}\}} \int_{\frac {d \mathcal{D}_{B}}{dt}>0} \frac {d \mathcal{D}_{B}(t, \rho_{\tau})}{dt} dt,
\end{equation}
where the trajectory states $\rho_{\tau}=\Lambda_{\tau}(\frac {\mathbf{I}}{d}), 0<\tau<\infty$ are the evolving state from the maximally mixed initial state $\rho_{0}=\frac {\mathbf{I}}{d}$ and the Bures angle is $\mathcal{D}_{B}(t, \rho_{\tau})=\mathcal{D}_{B}[\Lambda_{t}(\frac {\mathbf{I}}{d}), \Lambda_{t}(\rho_{\tau})]$.

In the case of the spin bath model, we can obtain the Bures angle between the maximally mixed state and the selectable trajectory state as,
\begin{equation}
\label{eq16}
\mathcal{D}_{B}=\arccos \frac 12\sum_{\theta=\pm 1} \sqrt{(1+\theta r_{3,t})(1+\theta r_{3,t}+\theta \mathcal{T}_{33,t}r_{3,\tau})}.
\end{equation}
Figure. 2 shows the effects of the local magnetic field $\epsilon$ and the bath temperature $T$ on the dynamics of the Bures angle. It is found out that the non-Markovian revivals of the Bures angle are mostly pronounced at the low bath temperature under the resonant condition of $\epsilon=2J$. On one hand, the non-Markovian behavior is suppressed when the detuning case of $\epsilon> 2J$ is considered. With the increasing of detuning, the interactions between the central qubit and bath are weaker than the transition energy of the qubit applied by the local field. Therefore, the impacts of the environment on the dynamics of the qubit become fragile. On the other hand, the high bath temperature dynamical process can also reduce the non-Markovian behavior of the Bures angle. This is the reason that the bath at high temperature appears to be in a very disorderly mixed state. The information of the qubit very quickly loses to the bath and very little information can return to the system from the environment. 

As one efficient application of the nonunital non-Markovian dynamics, it is of value to investigate the nonunital non-Markovian effects on the precision of quantum estimation. Quantum Fisher information \cite{Helstrom,Holevo} is a key quantity for describing the sensitivity of a quantum state with respect to a parameter $\chi$. Quantum Fisher information can provide a lower bound of the variance of any unbiased estimator due to the quantum Cra\'{m}er-Rao inequality \cite{BraunsteinPRL94}. A large value of quantum Fisher information represents an attainable measurement with a high precision. Among many versions of quantum Fisher information, there is a famous definition \cite{Helstrom,Holevo} as, 
\begin{equation}
\label{eq17}
\mathcal{F}_{\chi}:=\mathrm{Tr}(\frac {\partial \rho_{\chi}}{\partial \chi}\mathcal{L}_{\chi})=\mathrm{Tr}(\rho_{\chi}\mathcal{L}^{2}_{\chi}).
\end{equation}
The symmetric logarithmic derivatives  $\mathcal{L}_{\chi}$ is determined by $\frac {\partial \rho_{\chi}}{\partial \chi}=\frac 12\{\rho_{\chi} , \mathcal{L}_{\chi} \}$. By diagonalizing the density matrix for the quantum state as $\rho_{\chi}=\varrho_{m}|\psi_{m}\rangle \langle \psi_{m}|$ with $\sum_{m}\varrho_{m}=1$, we can transform the expression of the quantum Fisher information into,
\begin{equation}
\label{eq18}
\mathcal{F}_{\chi}:=\sum_{m}\frac {1}{\varrho_{m}}(\frac {\partial \varrho_{m}}{\partial \chi})^2+2\sum_{m \neq k}\frac {(\varrho_{m}-\varrho_{k})^2}{\varrho_{m}+\varrho_{k}}\big| \langle \psi_{m}| \frac {\partial}{\partial \chi}\psi_{k}\rangle\big|^2.
\end{equation}

The amplitudes $\vartheta$ for quantum states are changing because of the nonunital non-Markovian decocherence. We will evaluate the precision of quantum estimation about the amplitudes. We may assume that the initial composite state is chosen as the product state $|\Psi_{\vartheta}(0)\rangle=\Pi_{j=\mathrm{A,B}}^{\otimes}\cos \frac {\vartheta}{2}|1_{j}\rangle+\sin \frac {\vartheta}{2}|0_{j}\rangle$ with $\vartheta \in[0,\pi)$. According to Eq. (18), the quantum Fisher information with the parameter $\vartheta=\frac {\pi}{2}$  is calculated as,
\begin{equation}
\label{eq19}
\mathcal{F}^{\mathrm{Prod}}_{\vartheta=\frac {\pi}{2}}=\mathcal{T}_{33}^{2}+\frac { r_{3}^{2}\mathcal{T}_{33}^{2}}{1-r_{3}^{2}-\mathcal{T}_{11}^{2}-\mathcal{T}_{12}^{2}}.
\end{equation}
If the initial entangled state is given as $|\Psi_{\vartheta}(0)\rangle=\cos \frac {\vartheta}{2}|11\rangle_{\mathrm{AB}}+\sin \frac {\vartheta}{2}|00\rangle_{\mathrm{AB}}$, the quantum Fisher information $\mathcal{F}^{QC}_{\vartheta}$ with the parameter $\vartheta$ can also be numerically obtained by Eq. (18). The density matrix for the composite states can be diagonalized as $\rho^{\mathrm{A}\circ \mathrm{B}}(t)=\sum_{m=1}^{4}\varrho_{m}(t)|\psi_{m}\rangle \langle \psi_{m}|$ with the eigenvalues and corresponding eigenvectors as,
\begin{widetext}
\begin{eqnarray}
\label{eq20}
\varrho_{1,2}&=&\cos^{2}\frac {\vartheta}{2} \alpha \xi+\sin^{2}\frac {\vartheta}{2} \beta \delta,\nonumber \\
\varrho_{3,4}&=&\frac 12\cos^{2}\frac {\vartheta}{2}(\alpha^2+\xi^2) +\frac 12\sin^{2}\frac {\vartheta}{2}(\beta^2+\delta^2)\nonumber \\
& &+\frac 12\sqrt{\cos^{2}\frac {\vartheta}{2}\left[ (\alpha^2-\xi^2)+\sin^{2}\frac {\vartheta}{2}(\beta^2-\delta^2)\right]^2+\sin^{2}\vartheta \big| (\mathcal{T}_{11}+i\mathcal{T}_{12})^2\big|^2 },\nonumber \\
|\psi_{1,2}\rangle &=& |10\rangle_{\mathrm{AB}} (|01\rangle_{\mathrm{AB}}),\nonumber \\
|\psi_{3,4}\rangle&=&x_{3,4}|11\rangle_{\mathrm{AB}}+y_{3,4}|00\rangle_{\mathrm{AB}}.
\end{eqnarray}
\end{widetext}
Here, $\alpha=\sum_{n=0}^{\infty}A_{1}A_{1}^{\ast}e^{-\frac {2Jn}{T}}$, $\beta=\sum_{n=0}^{\infty}nC_{1}C_{1}^{\ast}e^{-\frac {2Jn}{T}}$ and $\xi=\alpha-\mathcal{T}_{33}-r_{3}, \quad \delta=\beta+\mathcal{T}_{33}-r_{3}$. The coefficients of the eigenvectors for $j=3,4$ are calculated by $x_{j}/y_{j}=\sin \vartheta(\mathcal{T}_{11}+i\mathcal{T}_{12})^2/[2(\varrho_{j}-\cos^{2}\frac {\vartheta}{2} \alpha^2-\sin^{2}\frac {\vartheta}{2} \beta^2)]$ and $|x_{j}|^{2}+|y_{j}|^{2}=1$.

The dynamical behavior of the quantum Fisher information for composite states with respect to the amplitude are shown in Figure 3. It is clearly seen that the oscillation of $\mathcal{F}_{\vartheta}^{QC}$ is markedly strong in the resonant case of $\epsilon=2J$. With the increasing of the bath temperature $T$, the revivals of $\mathcal{F}_{\vartheta}^{QC}$ are reduced. Under the condition of the large detuning $\epsilon=6J$, the values of the quantum Fisher information weakly oscillate in the vicinity of a certain high value. Besides it, the quantum Fisher information for composite entangled states $\mathcal{F}_{\vartheta}^{QC}$ is always superior to that for product state $\mathcal{F}_{\vartheta}^{Prod}$. It is found out that the nonunital non-Markovian effects can enhance the precision of quantum parameter estimation. 

To furthermore demonstrate the relations of quantum Fisher information and nonunital non-Markovianity, we study the derivatives of them with respect to time. The positive derivatives of quantum Fisher information denote the backflow of the information from the bath to system, which is regarded as non-Markovian behavior. Meanwhile, we use the Bures angle to evaluate the degree for nonunital non-Markovianity. In Figure 4, the dynamics of $\frac {d \mathcal{F}_{\vartheta}}{d t}$ is synchronous with that of $\frac {d \mathcal{D}_{B}}{d t}$. That is, the increase of the quantum Fisher information also represents the existence of the nonunital non-Markovian effects.

\section{discussion}

The nonunital non-Markovian effects from the environment can be studied by the time-dependent transfer matrix and inhomogeneity vector which are determined by the decoherence factors. The nonunital non-Markovian dynamics of the open system in a spin bath is analytically obtained in the thermodynamics limit by using the special operator method. 

We may select any two orthogonal initial states with collinear Bloch vectors. The nonunital non-Markovian environment leads to the two evolving states with non-collinear vectors. The revivals and suppressing of the angles between the two Bloch vectors happen in the nonunital non-Markovain dynamics, which is different from the monotonic decrease of  the angles in the Markovian dynamics. 

We also use the Bures angle to measure the degree for the nonunital non-Markovianity. In the resonant case, the nonunital non-Markovain effects are prominent at the low bath temperature. As one possible application, the nonunital non-Markovian bath can give rise to the enhancement of the precision of quantum parameter estimation. The increase of the quantum Fisher information for composite states with respect to the amplitude is in accordance with that of the Bures angle. This also provides another efficient way to the quantitative description of the nonunital non-Markovian effects.

\section{acknowledgment}

This work is supported by the National Natural Science Foundation of China under Grant No. 11074184, 11174363 and  No. 11174114. X. H. is financially supported from the China Postdoctoral Science Foundation funded project No. 2012M520494, the Basic Research Funds in Renmin University of China from the central government project No. 13XNLF03.

\newpage

{\Large \bf Figure Captions}

{\bf Fig. 1}

(a). The dynamical behavior of the two orthogonal with the collinear Bloch vectors during the time interval $Jt\in [0, 3]$ are plotted in the Bloch sphere when $\epsilon=2J$, $J_{0}=J$ and $T=J$; (b). The dynamics of these vectors during the time interval $\gamma t\in [0, 3]$ is also shown in the nonunital Markovian map when the decaying parameter is $p(t)=\exp(-\gamma t)$ with the positive value of $\gamma>0$. The black lines and blue ones denote the Bloch vectors for the initial state $|\psi_{1}(0)\rangle=\frac {1}{\sqrt 2}(| 1\rangle+| 0 \rangle)$ and $|\psi_{2}(0)\rangle=\frac {1}{\sqrt 2}(| 1\rangle-| 0 \rangle)$ respectively. The black solid arrows are the Bloch vectors at $t=0$ and blue dotted arrows represent the Bloch vectors after the time interval.

{\bf Fig. 2}

The dynamical processes of the Bures angle between the Bures angle between the maximally mixed state and the selectable trajectory state are plotted in order to measure the degree for the nonunital non-Markovianity under the condition of $\tau=2$ and $J_{0}=J$. 

{\bf Fig. 3}

The nonunital non-Markovian effects of the bath on the quantum Fisher information for composite states with the parameter $\vartheta=\pi/2$ are demonstrated when the bath temperature $T$ and the local magnetic field $\epsilon$ are changed. In the case of $\epsilon=6J$, $T=J$ and $J_{0}=J$, the blue solid line represents $\mathcal{F}^{QC}_{\vartheta}$ for the initial entangled state and green line denotes $\mathcal{F}^{Prod}_{\vartheta}$ for the product coherent state. The black dotted line is the one for $\mathcal{F}^{QC}_{\vartheta}$ when $\epsilon=2J$, $T=J$ and $J_{0}=J$. And the red dashed line is the one for $\mathcal{F}^{QC}_{\vartheta}$ when $\epsilon=2J$, $T=6J$ and $J_{0}=J$.

{\bf Fig. 4}

The time-dependent derivatives of the Bures angle (red line) and quantum Fisher information (black line) for entangled initial states with $\vartheta=\pi/2$ are shown when $\epsilon=6J$, $T=J$ and $J_{0}=J$.

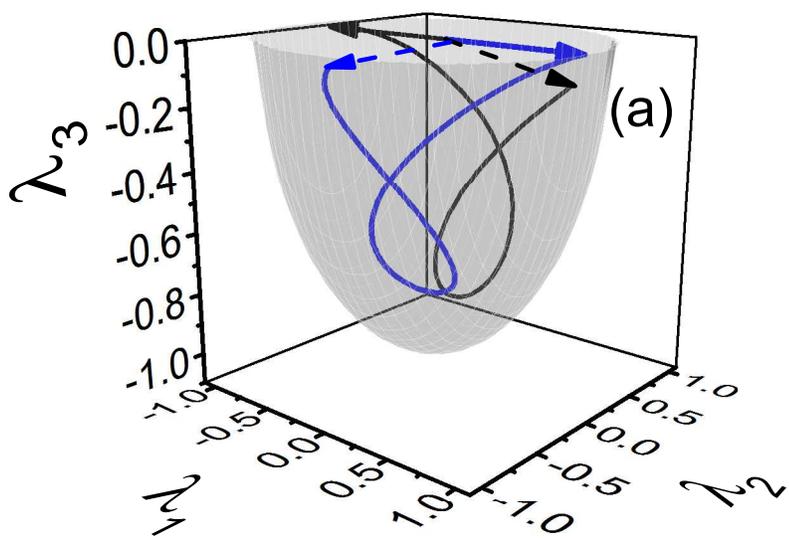
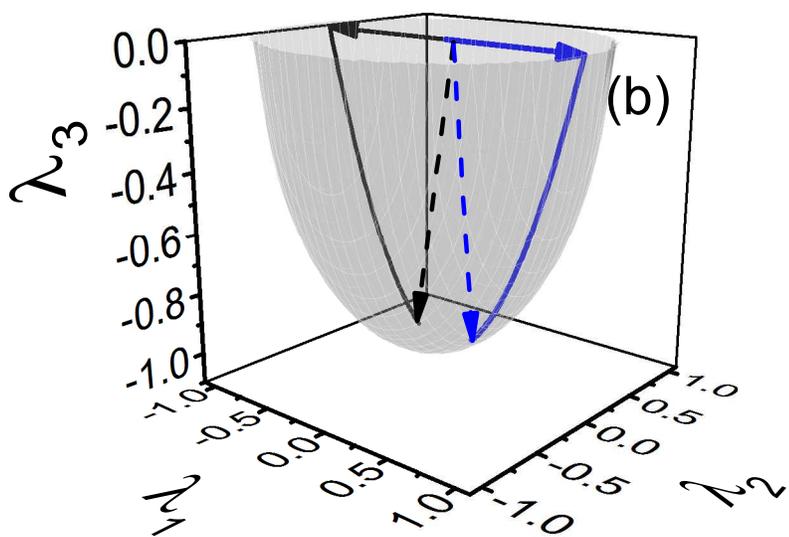
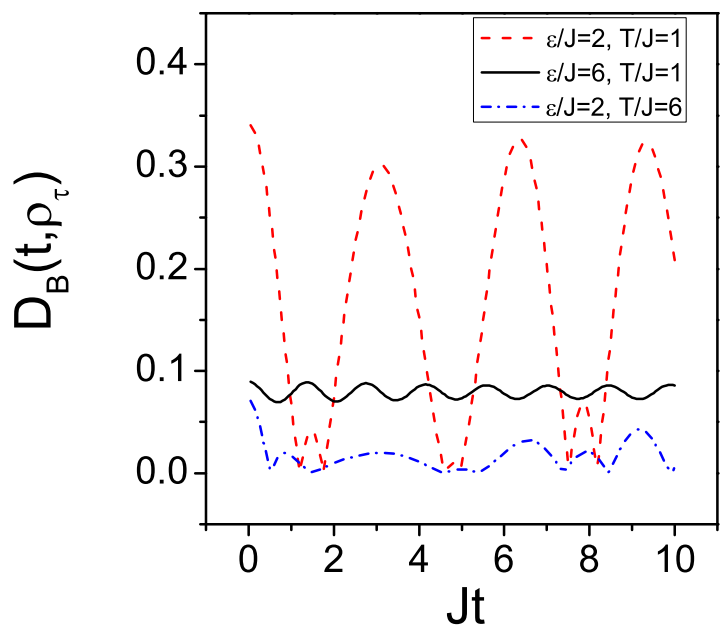
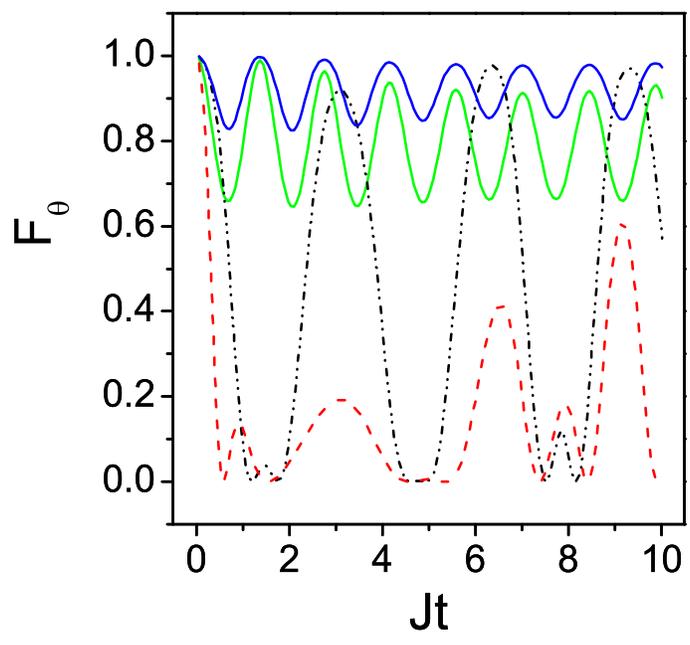
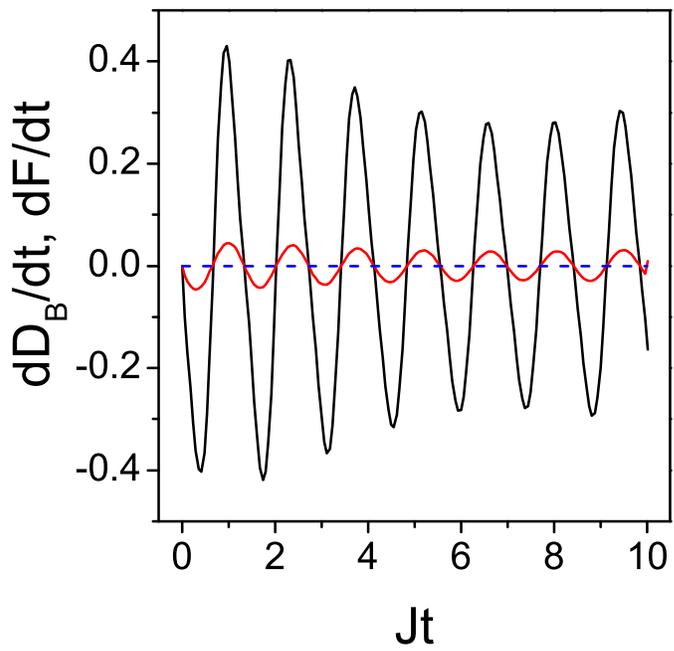
\end{document}